\documentstyle[11pt]{article}

\def\be{\begin{equation}}
\def\ee{\end{equation}}
\def\bea{\begin{eqnarray}}
\def\eea{\end{eqnarray}}

\begin{document}

\section*{\center  NEUTRINO OSCILLATIONS AT REACTORS: WHAT NEXT?}

\Large
{\hspace{3 cm} L.A. Mikaelyan, V.V. Sinev}
\\

\normalsize
{\it \center \quad RRC "Kurchatov Institute", Kurchatov Sq., 1, Moscow-123182, Russia.\\
 Talk given at the International Conference on Non-Accelerator New
Physics, NANP'99, Dubna, 28.06-3.07. 1999.}
\vspace{1em}\\
\righthyphenmin=3
 We shortly review past and future experiments at reactors aimed 
   at searches for neutrino masses and mixing. We also consider  
   new idea to search at Krasnoyarsk for small mixing angle 
   oscillations in the atmosheric neutrino mass parameter region.

\section{Introduction}
\large
 The first long baseline reactor experiment CHOOZ'97 [1] has 
success-fully reached the atmospheric  mass parameter region 
${\delta}m^{2}_{atm} \sim 10^{-3} eV^{2}$ and has tested there a large 
portion of the area of interest on the ${\delta}m^{2} - sin^{2}2{\theta}$ 
plane. No signs of oscillations have been 
found. Thus oscillations of electron neutrinos can NOT dominate 
in the atmospheric neutrino anomaly. 

 The Super-Kamiokande data on atmospheric neutrinos provide \\
strong evidence in favor of intensive 
${\nu}_{\mu} \rightarrow {\nu}_{X} \quad (x \ne e)$ 
transitions [2]. In the three active neutrino (${\nu}_{e}$, ${\nu}_{\mu}$ , ${\nu}_{\mu}$)
oscillation model considered here ${\nu}_{X} = {\nu}_{\tau}$.

We wish to emphasize however that both experiments, CHOOZ'97 and 
SK, do not exclude ${\nu}_{e} \leftrightarrow {\nu}_{\mu}$ oscillations as a 
subdominant mode in the ${\delta}m^{2}_{atm}$ region [3, 4].                                  

The results of recent experiments have attached great attention 
to the problem of neutrino oscillations. New physical ideas and 
projects of new large scale experiments at accellarators are actively 
discussed (for review cf. Ref.[4]).

What new contributions can be done with reactor ${\widetilde {\nu}}_{e},e$'s for 
exploring the electron neutrino mass and mixing problems?

One line of future studies has already been announced. To probe 
the large mixing angle (LMA) MSW solution (${\delta}m^{2}_{sol} \simeq 10^{-4}-10^{-5} eV^{2}$, 
$sin^{2}2{\theta} \sim 0.7$) [5] of the solar neutrino puzzle the projects KamLAND 
at Kamioka [6] and BOREXINO at Gran Sasso [7] plan to detect neutrinos from 
reactors which operate hundred kilometers away from the detector sites. 

In this report we consider another possibility.  We find 
that with two detector techniques the sensitivity to the mixing 
parameter in the ${\delta}m^{2}_{atm}$ region can be substantially increased in 
comparison with that achieved in CHOOZ experiment. We suggest a 
new study of the problem at Krasnoyarsk underground (600 m.w.e.)
laboratory with detectors stationed at  1100 m and  250 m from 
the reactor. The main goals of the proposed experiment are:

\begin{itemize}
\item To obtain better understanding of the role the electron 
 neutrino can play in the atmospheric neutrino anomaly,
\item To obtain new information on neutrino mixing, here the $U_{e3}$ \\
 element of the neutrino mixing matrix,
\item To give normalization for future long baseline experiments
 at accellarators.
\end{itemize}

\section{OSCILLATIONS OF REACTOR ANTI-\\NEUTRINOS}

Nuclear reactor generates antineutrinos at a rate $N_{\nu} \sim 1.8\cdot10^{20} s^{-1}$
per 1 GWt thermal power. The typical reactor ${\widetilde {\nu}}_{e}$ energy spectrum 
normalized to one fission event is presented in Fig.1. 

The ${\widetilde {\nu}}_{e}$s are detected via the inverse beta decay reaction

\begin{equation}
{\widetilde {\nu}}_{e} + p \rightarrow e^{+} + n , 
\end{equation}

The positron kinetic energy T is related to the energy E of the ${\widetilde {\nu}}_{e}$ as 
$$
T = E - 1.804 \quad MeV. \eqno (1')
$$

The signature that identifies the ${\widetilde {\nu}}_{e}$ absorption in the liquid 
scintil-lator target is a spatially correlated delayed coincidence 
between the prompt positron and the signal from the neutron capture 
${\gamma}$-rays. 

The probability P(${\widetilde {\nu}}_{e}\rightarrow {\widetilde {\nu}}_{e}$) for 
${\widetilde {\nu}}_{e}$ to survive at the distance of 
R meters  from the source is given by the expression 

\begin{equation}
P({\widetilde {\nu}}_{e}\rightarrow {\widetilde {\nu}}_{e}) = 1 - sin^{2}2{\theta}\cdot sin^{2}(1.27{\delta}m^{2}\cdot R\cdot E^{-1}),
\end{equation}

E(MeV) is the neutrino energy, ${\delta}m^{2}$ is the mass parameter in $eV^{2}$ and $sin^{2}2{\theta}$ 
is the mixing parameter.

The distortion of the positron energy spectrum and the deficite 
of the total ${\widetilde {\nu}}_{e}$ detection rate relative to no-oscillation 
case  are the signatures for oscillations that are searched for in the 
experiment. The deficite of the total rate is the strongest for 
$(R\cdot{\delta}m^{2})_{max} \approx 5 \ m\cdot eV^{2}$.
          
In the pressurized water reactors (PWR) ${\widetilde {\nu}}_{e}$ spectrum and 
total cross section of the reaction (1) vary with the nuclear fuel composition,
(the burn up effect). The current fuel composition is provided by 
reactor services. When the fuel composition is known the 
no-oscillation cross section ${\sigma}_{V-A}$ can be calculated within the 
uncertainty of 2.7\%. (For more information see e.g.[8] and references 
therein).

With the help of an integral type detector the CdF-KURCHATOV-LAPP 
group has accurately measured the cross section at a 15 m distance 
from the Bugey-5 reactor [9]:

\begin{equation}
{\sigma}_{exper} = 5.750\cdot 10^{-43} cm^{2}/fission \pm 1.4\%
\end{equation}

This highly accurate value ${\sigma}_{exper}$ can be used in other 
experiments with reactor ${\widetilde {\nu}}_{e}$'s as a no-oscillation 
metrological standard. When it is used in practice one must consider the 
differences in the fuel composi-tions and take into account a number of 
"small effects". This increases the error up to about 2\%. 

\section{PAST, CURRENT AND FUTURE \\ EXPERIMENTS}

During 1980 - 1995 yy. intensive searches for neutrino oscillations 
with detectors located at $\sim$ 10 - 230 m from relevant reactors have 
been performed. These "short baseline" experiments are listed in the 
left part of Fig.2. The highest sensitivity to the mixing parameter 
($sin^{2}2{\theta} \approx 0.02$) was reached by the Bugey-3 group in the 
measurements with two identical detectors located at 15 m and 40 m from the 
reactor [9] (see Fig.3).

The CHOOZ detector used a 5 tons liquid scintillator (Gd) target.
It was located in an underground laboratory (300 MWE) at a distance of 
about 1 km from the neutrino source. The ratio R of the measured to
expected in no oscillation case neutrino detection rates was  
\begin{equation}
R = 0.98 + 0.04 (stat) + 0.04 (syst), /November \quad 1997/.
\end{equation}
The systematic errors come mainly from the reactor properties and 
the absolute values of neutrino detection efficiencies. The 90\% CL 
exclusion plot "CHOOZ'97" for ${\widetilde {\nu}}_{e}$ disappearance channel 
is presented in Fig.3. together with the allowed ${\nu}_{\mu} \rightarrow {\nu}_{\tau}$  oscillation channel 
SK'736 d [2] (shaded area). The experiment was continued till 
June 1998 y. to achieve better statistics and improve systematics.
The final CHOOZ results will appear soon.

The Palo Verde oscillation experiment at $\sim$ 800 m from three 
reactors is taking data since October 1998. The first 70 day results 
are now available [11]. 

Past and current experiments cover now the distances up to 1 km 
from the reactor. The extension to $\sim$ 200 and $\sim$ 800 km is expected
with the forthcoming KamLAND and BOREXINO ultra long baseline projects 
(Fig.2). They will use liquid scintillator targets of 1000 and 300 
tons respectively. The solar large mixing angle MSW solution [7] is 
well inside the area planned for the investigation (Fig.3).

The experimental goal of the new search at Krasnoyarsk is to extend 
studies to the "white spot" area left to the CHOOZ limits in Fig.3.

\section{NEW PROJECT FOR KRASNOYARSK}

\subsection{Detectors}

Two identical liquid  scintillation spectrometers stationed at the 
Krasno-yarsk underground site (600 MWE)  at  distances $R_{1}$ = 1100 m  
and $R_{2}$ = 250 m from the reactor source simultaneously detect ($e^{+}$,n)
pairs produced via reaction (1). A simplified version of the BOREXINO 
detector composition is chosen for the design of the spectrometers 
(Fig.4). The 50 ton targets in the center of the detectors 
(mineral oil + PPO) are viewed by the PMT's ($\sim 20\%$ coverage, $\sim$ 120 
ph.e./MeV) through $\sim$ 1 m thick layer of non-scintillating oil. 

Computed neutrino detection rates can be seen in the middle of
Table 1. For comparison parameters of the CHOOZ and future 
Kam-LAND and BOREXINO detectors is also included. 

\begin{table}[t]
\caption{Antineutrino detection rates $N(e^{+}$,n) $d^{-1}$.\label{tab:exp1}}
\vspace{0.4cm}
\begin{center}
\begin{tabular}{|c|c|c|c|c|}
\hline
Detector & {\small CHOOZ'97} & {\small THIS PROJECT} & KamLand & {\small BOREXINO} \\
\hline
 Mass of the  &  &  &  & \\
 target, tons & 5 & 50 \qquad 50 & 1000 & 300 \\
 Distance from & & & & \\
 the source, km  & 1 & 0.25 \quad 1.1 & $\sim$ 200 & $\sim$ 800 \\
 $N(e^{+}$,n) $d^{-1}$ & 12 & 1000 \quad 55 & 2 & 0.08 \\
\hline
\end{tabular}
\end{center}
\end{table}

\subsection{Backgrounds}

The CHOOZ experiment has demonstrated revolutionary improvements
of the reactor neutrino techniques: a 500 - 1000 times lower level the 
background has been achieved relative to previous experiments at 
reactors (see the first three colomns in Tab.2).

It is important to note that with  the CHOOZ experience and also 
with detailed studies performed at the BOREXINO CTF detector [7] the 
main secrets of the background supression are now well understood, at 
least for the level we need here. The secrets are:

\begin{itemize}
\item Detector should be  stationed sufficiently deep underground
to reduce the the flux of cosmic muons, the main source of the 
background in these type experiments. 
\item To reduce the accidental background the potomultiplyers with 
their highly radioactive glass should be separated from the 
central scintillator volume by sufficiently thick layer of the 
oil ("Borexino geometry", Fig.4). 
\end{itemize}

We estimate the total background rate as 0.1 per day, per ton 
of the target. It is 2.5 times lower than the background measured 
at CHOOZ, which seems reasonable for a detector located twice as 
deep underground (compare Tab.2).

\begin{table}[t]
\caption{Neutrino signal $N(e^{+}$,n) and background $N_{BKG}$ rates
(per 1 day, per 1 ton of scintillator target).
\label{tab:exp2}}
\vspace{0.4cm}
\begin{center}
\begin{tabular}{|c|c|c|c|c|c|c|}
\hline
{\small Detector} & {\small Rovno} & {\small $Bugey^{*}$} & {\small CHOOZ'97} 
& {\small This} & {\small Kam} & {\small BOREXINO} \\
 &  &  &  & {\small $Project^{**}$} & {\small LAND} & \\
\hline
 {\small $MWE^{***}$} & 30 &  $\sim$ 10 & 300 &  600  &  2700 &  3200 \\
 {\small $N(e^{+}$,n)} & 1700 & 370 & 2.4 & 1.1  & $2\cdot 10^{-3}$ & $3\cdot 10^{-4}$ \\
 {\small $N_{BKG}$} & 220 & 160 & 0.24 & $\sim$ 0.1 & $<$ 10 & $<$ 10 \\
\hline
\multicolumn{7}{l}{$^{*}$ \small Detector at 40 m position.} \\   
\multicolumn{7}{l}{$^{**}$ \small Detector at 1100 m position.} \\
\multicolumn{7}{l}{$^{***}$ \small Overburden in meters of water equivalent.} \\
\end{tabular}
\end{center}
\end{table}

\subsection{Data analysis}

In three years of data taking $40\cdot 10^{3}$  neutrino events with 10:1 
the signal to background ratio and about $800\cdot 10^{3}$ events can be 
accumulated at the 1100 m 250 m positions respectively.

Two types of analysis can be used.  Both of them are not affected 
by the value of the absolute ${\widetilde {\nu}}_{e}$ flux and 
${\widetilde {\nu}}_{e}$ energy spectrum, reactor 
power, the burn up effects and absolute values of the detector 
efficiencies. 

Analysis I is based on the ratio $X_{RATE}=N_{1}/N_{2}$ of the neutrino 
detection rates measured at two distances:

\begin{equation}
 X_{RATE}=\frac{R^{2}_{2}\cdot {\epsilon}_{1}\cdot V_{1}}{R^{2}_{1}\cdot {\epsilon}_{2}\cdot V_{2}}\cdot F({\delta}m^{2},sin^{2}2{\theta})
\end{equation}
${\epsilon}_{1,2}$ and $V_{1,2}$ are the neutrino detection efficiencies and the 
scintillator volumes. 
   
Thus the absolute values of detection efficiencies are practically
conceled, only their small relative differences are to be considered
here.

Analysys II is based on the comparison of the shapes of the $e^{+}$
spectra $S(E_{e}$ ) measured simultaneously in two detectors. Small 
devia-tions of the raio $X_{SHAPE}=S_{1}/S_{2}$

\begin{equation}
 X_{SHAPE}= C\cdot (1 - sin^{2}2{\theta}sin^{2}{\phi}_{1})\cdot (1 - sin^{2}2{\theta}sin^{2}{\phi}_{2})^{-1},
\end{equation}
from a constant value are searched for the oscillation effects 
(${\phi}_{1,2}$ stands for $1.27{\delta}m^{2}R_{1,2}E^{-1}$). No knowleadge of the constant C in 
Eq. (6) is needed for this analysis so that the details of the 
geometry, ratio of the target volumes and efficiencies are excluded 
from the consideration.  

\subsection{Detector calibrations}

Calibrations of the detectors are of crucial importance. 
Difference between the respons functions of the two detectors which 
is difficult to avoid can produce some modulation of the ratio (6) 
and thus imitate the oscillation effect. The differences can be 
measured and relevant corrections found. This can be done by 
systematic intercomparison of the scales in many energy points using 
the sources of $\gamma$-rays shown in Fig.5. 

An additional approach is also considerd. The spectrometers can 
be tested periodically with the use of the $^{252}Cf$ spontaneous fission 
source that can produce a broad spectrum due to prompt $\gamma$-rays and 
neutron recoils (Fig.5). The ratio of these spectra should be 
constant, if the instrumental modulation is observed it can be 
measured and used to find corrections to the Eq.(6). 

\subsection{Expected constraints on oscillation parameters}

We hope that the ratio ${\epsilon}_{1}V_{1}/{\epsilon}_{2}V_{2} \approx 1$ (Eq.5) 
can be controlled within 0.8\%. Then from Analysis I we expect the 90\% CL limits 
shown in Fig.6 (the curve labelled "RATE").

We believe that the false effects in Eq.(6) can be controlled down 
to the level of 0.5\%. Relevant 90\% CL limits are presented in Fig.6
(the curve "SHAPE").

\section{DISCUSSION AND CONCLUSIONS}

With methods of data analysis mentioned in Sec.4 we arrive at 
the limits on the oscillation parametes decoupled from the main 
sources of systematic uncertainties which limit the sensitivity of the 
experiments based on absolute comparison of the measured and expected 
no-oscilla-tion rates and/or positron spectra.

Nevertheless the systematic errors which remain notably decrease 
the sensitivity to the mixing parameter $sin^{2}2{\theta}$. The curve "SHAPE" 
(Fig.6) is about two times less restrictive as compared to the 
statistical limits found for an ideal detector with no systematics. 

Now we come back to the main question: what contributions to the 
neutrino physics can be expected from new oscillation experiments at 
reactors?

The long baseline (LBL) experiments with detectors stationed at a 
distance of $\sim$ 1 km from the reactor search for the mixing parameter 
$sin^{2}2{\theta}_{LBL}$ which is expressed in this case as 
\begin{equation}
 sin^{2}2{\theta}_{LBL}=4\cdot U^{2}_{e3}(1-U^{2}_{e3}),
\end{equation}
$U^{2}_{e3}$ is the contribution of the heaviest mass eigenstate ${\nu}_{3}$ 
to the flavor electron neutrino state:
\begin{equation}
 {\nu}_{e}=U_{e1}{\nu}_{1}+U_{e2}{\nu}_{2}+U_{e3}{\nu}_{3}
\end{equation}
From CHOOZ'97 results we already know that $U^{2}_{e3}$ is not large: 
$U^{2}_{e3} < (3 - 5)\cdot 10^{-2}$. Future experiment at 1 km considerd here 
can find $U^{2}_{e3}$ or set a much smaller upper limit. Therefore better 
understanding of the neutrino mixing can be achieved. New information on $U_{e3}$ 
can be useful for analysis of the atmospheric neutrinos and can give hints 
for future long baseline experiments at accelerators.

The ultra long baseline (ULBL) experiments KamLAND and BORE-XINO 
will search for $sin^{2}2{\theta}_{ULBL}$ which depends on the contributions 
of the ${\nu}_{1}$ and ${\nu}_{2}$ mass states:
\begin{equation}
 sin^{2}2{\theta}_{ULBL}=4\cdot U^{2}_{e1}U^{2}_{e2}
\end{equation}

We conclude that the experiments at reactors discussed here can 
provide full information of the mass structure of the electron 
neutrino, at least in the three neutrino oscillation model.

\section*{Acknowledgments}

We greatly appreciate fruitful discussions with S.Bilenky, E. Lisi 
and A. Smirnov. We thank our colleagues V. Martemyanov, Yu. Kozlov 
and V. Vyrodov for many discussions. This study is supported by RFBR.


\end{document}